\documentstyle[epsf]{article}
\begin{document}
\begin{center}
{\bf MSSM HIGGS SECTOR AT THE ONE-LOOP LEVEL}\footnote[1]{to be published in Czech Journal of Physics}
\vskip 8mm
{\sc M. Malinsk\'{y}}\footnote[2]{malinsky@hp02.troja.mff.cuni.cz}\\
{\small \it Department of Particle and Nuclear Physics,\\ 
Faculty of Mathematics and Physics, Charles University, \\ V Hole\v{s}ovi\v{c}k\'{a}ch 2, 180 00 Praha 8, Czech Republic}
\end{center}
\renewcommand{\cos}[2]{{\rm cos}^{#2}{#1}\,}
\renewcommand{\sin}[2]{{\rm sin}^{#2}{#1}\,}
\renewcommand{\tan}[2]{{\rm tan}^{#2}{#1}\,}
\renewcommand{\cot}[2]{{\rm cot}^{#2}{#1}\,}
\newcommand{\mln}[1]{\ln{\frac{#1}{\mu^2}}}
\newcommand{\lag}[1]{{\cal L}_{\sc #1}}
\newcommand{\tr}{{\rm \,Tr}}
\newcommand{\beq}{\begin{equation}}
\newcommand{\eeq}{\end{equation}}
\newcommand{\beqa}{\begin{eqnarray}}
\newcommand{\eeqa}{\end{eqnarray}}
\newcommand{\sto}{{\tilde{t}_1}}
\newcommand{\stt}{{\tilde{t}_2}}
\newcommand{\stl}{{\tilde{t}_L}}
\newcommand{\str}{{\tilde{t}_R}}
\renewcommand{\tt}{\theta_t}  
\newcommand{\tW}{\theta_W}  
\renewcommand{\div}{{\rm C_{\sc uv}}}
\newcommand{\dreq}{\stackrel{{\sc dr}}{\sim}}
\newcommand{\gmn}{g^{\mu\nu}}
\newcommand{\rel}{m_h^2+m_H^2-m_A^2-m_Z^2=0}
\newcommand{\s}[1]{#1 \!\!\!\slash}
\newcommand{\vare}[0]{\varepsilon}
\newcommand{\intf}[3]
{\int\!\!\frac{d^4 #1}{(2\pi)^4}\frac{#2}{#3}}
\newcommand{\intfp}[2]
{\int\!\!\frac{d^4 #1}{(2\pi)^4} #2 }
\newcommand{\intdp}[2]
{\int\!\!\frac{{\rm d}^d #1}{(2\pi)^d} #2 }
\newcommand{\intx}
{\int_0^1\!\!{\rm d}x}
\newcommand{\intt}
{\int\!\!{\rm d}^4\theta}
%
\newcommand{\intd}[3]
{\int\!\!\frac{d^d #1}{(2\pi)^d}\frac{#2}{#3}}


\begin{abstract}
This work provides an elementary introduction to the Higgs sector renormalisation 
within the Minimal Suppersymmetric 
Standard Model (MSSM) framework. The main aim of the paper is to clarify some technical 
details that are usually omitted in the existing literature. 
The MSSM tree-level relation $m_h^2+m_H^2=m_A^2+m_Z^2$ is renormalised using the
standard technique of direct computation of the relevant one-loop Feynman
diagrams.
The calculation is performed within the unitary gauge
and the definition of the renormalised parameters is briefly reviewed.
The expected cancellation of ultraviolet divergences is explicitly checked and the 
well-known leading-log 
term is recovered. All the necessary ingredients of the computations are sumarized 
in the appendices which 
makes the work more self-contained.   
\end{abstract}

\section{Introduction}     

The Minimal Supersymmetric Standard Model (MSSM) provides one of
the first realistic attempts to describe physics beyond the Standard Model.
Although there is a huge amount of new fields and parameters in
any SUSY-extended model, the MSSM is able to reproduce all the successful
predictions of the
original GWS theory with a very good accuracy. 
The MSSM Higgs sector is enlarged due to the fact that supersymmetry forbids the usual
mechanism of generation of  the up-type quark masses. This problem requires introduction 
of another Higgs doublet with opposite charges, which produces  all the necessary quark
mass-terms. Consequently there are five Higgs particles and three unphysical 
Goldstone bosons  in the theory. 

The global $N=1$ supersymmetry, being a new symmetry in addition to the original
$SU(3)_c\otimes SU(2)_{\sc l}\otimes U(1)$ local gauge invariance,
puts constraints on the Higgs sector, which is
rather indefinite in nonsupersymmetric theories. In the case
of MSSM this leads to the famous tree-level relation $m_h\leq
m_Z|\cos{2\beta}{}|\quad$\cite{GaH},\cite{Sim} which bounds
the mass of the lightest Higgs scalar. Note that up to now there is no experimental evidence 
of such state; the present day lower limit is $m_h\geq 88.3 \, GeV,$  see \cite{Last},\cite{Ep}.
However, radiative corrections modify the upper bound
substantially \cite{Ber},\cite{YaY}.
This bound is closely related to the tree-level sum-rule 
\beq
\label{rel1}
\rel
\eeq
Renormalising this rule one can obtain the shift of the unsatisfactory bound descending
from the radiative corrections.
Up to now there are many papers performing the computation on the  one-lop level 
\cite{Ber},\cite{YaY},\cite{GaT},\cite{Haber} and
also some  two-loop results were already obtained \cite{HHW}.
Most of these papers are concentrated on the evaluation of the finite
part of the correction only without explicit discussion of the divergences. It is 
expected that the ultraviolet divergences in such relations originating from the additional 
symmetry cancel, but  there is 
still no general proof of this based on the relevant Ward identities. From this 
point of view it may be useful to demonstrate explicitly the mechanism of the compensation
in the particular case of relation (\ref{rel1}); this, in fact,  is one of the goals of this work.
In this respect, this paper supplements the calculations presented in the cited 
literature.

The whole analysis is performed for the top-stop sector only. 
It is sufficient due to the observation that the leading term coming from any 
fermion-sfermion cluster is proportional to the fourth power of the fermionic mass  
\cite{Ber},\cite{YaY}. Note that the contributions coming from the chargino and neutralino
sectors are negligible \cite{GaT}.   

The paper is organized  as follows: the definition of the renormalised
parameters is briefly reviewed in section 1; the U-gauge
allows us to simplify the matter essentially compared to the
choice of \cite{Ber}. Section 2 is devoted to discussion of
the UV-divergences originating from the one-loop Feynman graphs
renormalising the relevant 2-point Green functions. The leading logarithmic
term is recovered in the third section. Most of the technical details are deferred to 
Appendices.
%
%
%
%
%
%
%
%
%
%
%
%
%
\section{Definition of the renormalised parameters}

As was stated before the whole computation will be performed in
the unitary gauge; this particular choice reduces the number of
diagrams to be considered and simplifies the renormalisation
scheme. 
On the other hand, the presence of Goldstone bosons in $R_{\xi}$ gauges causes
for example the
total cancellation of contributions descending from the tadpole
diagrams (see \cite{Ber}), which does not occur in U-gauge.

The diagrams to be considered
are listed in Appendix A. The ultraviolet divergences coming from the loops
are handled using the standard technique of dimensional
regularisation, see \cite{Aok}.


\subsection{Renormalized (pseudo)scalar masses:}
\mbox{}
\newline
The renormalized masses $m_X$of (pseudo)scalars $h$,$H$ and $A$ are
defined generically by
\beq
\label{defm2}
\left(1+\delta Z_X\right) m_{XB}^2 =m_X^2+\delta m_X^2 ;
\eeq
  %
%
Next, let us denote the sum of all (relevant)  graphs by
$-i\Pi_X(q)=\sum -i\Pi^{Y}_X(q)$.
Then the 2-point Green functions can be written in
the form (see \cite{Bailin}).
$$
i\Gamma^{(2)}_X(q,-q)=q^2-m_X^2-i\Pi_X(q)+i\left(\delta Z_X q^2 - \delta m_X^2\right)
+{\rm higher}\>\>{\rm order}
$$
We adopt the so-called {\it on-shell} renormalisation scheme in which
all the external momenta ($q$'s)
are taken to be on the mass-shell, i.e.
$q^2=m^2_X$ where $m^2_X$ denotes the squared mass of the
considered particle.
In this scheme we use the following renormalisation conditions
$$
\Gamma^{(2)}_X(q,-q)=0 \quad, \qquad
\frac{\partial \Gamma^{(2)}_X}{\partial q^2}(q,-q)=1 \qquad {\rm at}
\quad q^2=m_X^2
$$
This particular choice implies
\beq
\delta Z_X = 0 +{\rm higher}\>\>{\rm order}\>; \qquad
\delta m_X^2=-\Pi_X\left(q^2=m^2_X\right)+{\rm higher}\>\>{\rm order}
\label{delm}
\eeq
The physical mass $m_X$ can then be expressed (using
(\ref{defm2}) and (\ref{delm})) as
$$
m_X^2=m_{BX}^2+\Pi_X(q^2=m_X^2)+{\rm higher}\>\>{\rm order}
$$


\subsection{Renormalized $Z$-boson mass:} \mbox{}
\newline
Let us denote the sum of all the one-loop ($Z$) 'vacuum polarisation
graphs' by $-i\Pi_Z^{\mu\nu}(q)$. This quantity renormalizes
the $Z$-boson mass to the new value


$$
m_Z^2=m_{ZB}^2-A_Z(q^2=m_Z^2) +{\rm higher}\>\>{\rm order}
$$
(we have again used the on-shell conditions) where $A_Z(q^2)$ is defined
by
$$\Pi_Z^{\mu\nu}(q^2)\equiv A_Z(q^2)\gmn+B_Z(q^2)q^\mu q^\nu$$
(i.e. corresponds to
the coefficient of the transverse part of $\Pi^{\mu\nu}_Z$).


\subsection{Renormalised sum-rule:}
\mbox{}\\
Having defined renormalised quantities we can recast the relation
(\ref{rel1}) in the renormalised form
$$
m_h^2+m_H^2-m_A^2-m_Z^2=\Delta+{\rm higher}\>\>{\rm order}
$$
where
\beq
\label{correction}
\Delta \equiv \Pi_h(q^2=m_h^2)+\Pi_H(q^2=m_H^2)-\Pi_A(q^2=m_A^2)+A_Z(q^2=m_Z^2)
\eeq
This is the most important relation of this section.
In the following part we attempt to evaluate the one-loop leading-log term of $\Delta$. 
As was already stated before the leading term descends from the graphs involving top and 
supertop loops so the rest of this computation will be performed for this sector only. 
Note that the full quantity includes contributions from almost all the particles in the theory, 
which would complicate the calculation essentially without any impact on the leading term so the 
other contributions are simply omitted.  
%
%
%
%
%
%
%
%
%
%
%
%
%
%
%
%
%
%
%
%
%
\section{\label{kompenzace}Cancellation of UV-divergences}
In this section we show that the UV-divergent parts of the diagrams
listed in Appendix A cancel. To proceed we  put the external momenta on-shell and
substitute in (\ref{correction}). 
For the sake of brevity
there will be no difference between the symbols used for the divergent parts of the
considered expressions and the full contributions in this section; moreover the overall
factors $\div$ and $N_c$ are suppressed too. For example $B_h^t$ (see Appendix A)
corresponds here to $g_{htt}^2(4\pi^2)^{-1}\left(3m_t^2-\frac{1}{2}m_h^2\right)$.
To simplify the reader's insight the definitions of the partial sums of divergences 
(denoted by UV with relevant sub- and superscripts) take care of the sign of the 
corresponding expressions in (\ref{correction}).
 
\subsection{UV divergences in graphs involving top loops}

Let us start with the divergences descending from the graphs involving
one top-quark loop. The first three graphs in (\ref{topsector}) give
\beq
B_h^t+B_H^t-B_A^t=
\label{p1}
\eeq
$$
=\frac{g^2m_t^2}{16\pi^2m_W^2\sin{\beta}{2}}
\left[3m_t^2-m_t^2\cos{\beta}{2}+\frac{1}{2}\left(-m_h^2\cos{\alpha}{2}
-m_H^2\sin{\alpha}{2}+m_A^2\cos{\beta}{2}\right)\right]
$$
Next, the divergent part of the fourth graph in (\ref{topsector}) contributing to (\ref{correction}) is
after some algebra
$$
B_Z^t=-\frac{g^2m_t^2m_Z^2}{32\pi^2m_W^2}
+\frac{g^2m_Z^4}{24\pi^2m_W^2}
\left({\vare^t_L}^2+{\vare^t_R}^2\right)
$$
Utilising relations(\ref{hmotyhiggsu}) of Appendix B  the tadpole graphs in  
(\ref{topsector}) give
\beq
\label{p3}-T_A^{ht}-T_A^{Ht}=-\frac{g^2m_t^4}{16\pi^2m_W^2}
\eeq
Summing up the partial results  (\ref{p1})-(\ref{p3}) one obtains the total divergence 
coming from the graphs involving one top-quark loop:
\beq
\label{0}{\rm UV}_{\sc top}=\frac{g^2}{16\pi^2m_W^2\sin{\beta}{2}}\left[
2m_t^4-m_t^2m_Z^2\sin{\beta}{2}+\frac{2}{3}
m_Z^4\sin{\beta}{2}\left({\vare^t_L}^2+{\vare^t_R}^2\right)\right]
\eeq

\subsection{UV divergences in graphs involving supertop loops}
The same brief list of divergences will be now built up for the diagrams with one
supertop loop. The first type graphs in (\ref{stopsector}) give
$$
{\rm UV}_{\sc stop}^{(1)}\equiv
B_h^{\tilde{t}_1\tilde{t}_1}+B_h^{\tilde{t}_2\tilde{t}_2}+B_h^{\tilde{t}_1\tilde{t}_2}+B_H
^{\tilde{t}_1\tilde{t}_1}+B_H^{\tilde{t}_2\tilde{t}_2}+B_H^{\tilde{t}_1\tilde{t}_2}-B_A^{\
tilde{t}_1\tilde{t}_2}=
$$
$$
=
\frac{1}{16\pi^2}\left(g^2_{h\sto\sto}+g^2_{h\stt\stt}+2g^2_{h\sto\stt}+
g^2_{H\sto\sto}+g^2_{H\stt\stt}+2g^2_{H\stt\stt}+
2g^2_{A\sto\stt}\right)
$$
The result of the computation is
$$
{\rm UV}_{\sc stop}^{(1)}=
\frac{g^2}{16\pi^2m_W^2\sin{\beta}{2}}\Bigl[-2m_t^4-\frac{1}{2}
m_t^2(A_tm_6\sin{\beta}{}+\mu\cos{\beta}{})^2+
$$
\beq
\label{1}+m_t^2m_Z^2\sin{\beta}{2}
-m_Z^4\sin{\beta}{2}\left({\vare^t_L}^2+{\vare^t_R}^2\right)\Bigr]
\eeq
Next, the total contribution descending from the second type graphs in (\ref{stopsector}) reads
\beq
\label{2}{\rm UV}_{\sc stop}^{(2)}\equiv B_Z^{\tilde{t}_1\tilde{t}_1}+ 
B_Z^{\tilde{t}_2\tilde{t}_2}+ B_Z^{\tilde{t}_1\tilde{t}_2}=
\frac{g^2m_Z^2}{16\pi^2m_W^2}\Bigl\{\frac{1}{3}m_Z^2
\left({\vare^t_L}^2+{\vare^t_R}^2\right)
-
\eeq
$$
-2\left[m_\sto^2\left({\vare^t_L}^2\cos{\tt}{2}+{\vare^t_R}^2\sin{\tt}{2}\right)
+m_\stt^2\left({\vare^t_L}^2\sin{\tt}{2}+{\vare^t_R}^2\cos{\tt}{2}\right)\right]\Bigr\}
$$
The divergence coming from the tadpole sector of (\ref{stopsector}) can be written in 
the form (note that the minus sign corresponds to the sign of $\Pi_A$ in 
(\ref{correction}))
$$
{\rm UV}_{\sc stop}^{(3)}\equiv 
-T_A^{h\tilde{t}_1}-T_A^{h\tilde{t}_2}-T_A^{H\tilde{t}_1}-T_A^{H\tilde{t}_2}
$$
with the result
\beq
{\rm UV}_{\sc stop}^{(3)}=
\label{3}
\frac{g^2}{32\pi^2m_W^2}\Bigl\{m_t^2\left(m_\sto^2+m_\stt^2\right)
+m_t^2\left(A_tm_6+\mu\cot{\beta}{}\right)^2+
\eeq
$$
+m_Z^2\cos{2\beta}{}
\left[m_\sto^2\left({\vare^t_L}\cos{\tt}{2}-{\vare^t_R}\sin{\tt}{2}\right)
+m_\stt^2\left({\vare^t_L}\sin{\tt}{2}-{\vare^t_R}\cos{\tt}{2}\right)\right]\Bigr\}
$$
The last part of the total UV divergence originates from the seagull-type diagrams in 
(\ref{stopsector}):
$$
{\rm UV}_{\sc stop}^{(4)}\equiv 
S_h^{\tilde{t}_1}+S_h^{\tilde{t}_2}+S_H^{\tilde{t}_1}
+S_H^{\tilde{t}_2}-S_A^{\tilde{t}_
1}-S_A^{\tilde{t}_2}+S_Z^{\tilde{t}_1}+S_Z^{\tilde{t}_2}
$$
After some algebra one gets
$$
{\rm UV}_{\sc stop}^{(4)}=
-\frac{g^2m_t^2}{32\pi^2m_W^2}\Bigl\{
\left(m_\sto^2+m_\stt^2\right)+
$$
$$
+m_Z^2\cos{2\beta}{}
\left[
m_\sto^2 \left({\vare^t_L}\cos{\tt}{2}-{\vare^t_R}\sin{\tt}{2}\right)+
m_\stt^2 \left({\vare^t_L}\sin{\tt}{2}-{\vare^t_R}\cos{\tt}{2}\right)
\right]-
$$
\beq
\label{4}
-4m_Z^2\left[
m_\sto^2\left({\vare^t_L}^2\cos{\tt}{2}+{\vare^t_R}^2\sin{\tt}{2}\right)+
m_\stt^2\left({\vare^t_L}^2\sin{\tt}{2}+{\vare^t_R}^2\cos{\tt}{2}\right)
\right]\Bigr\}
\eeq
With all the partial results (\ref{0}),(\ref{1}),(\ref{2}),(\ref{3}) and (\ref{4}) at hand it is 
already easy to state (utilising (\ref{hmotyhiggsu}) from Appendix B) that 
the divergent parts of the considered diagrams contributing to the relation 
(\ref{correction}) exactly cancel :
$$
{\rm UV}_{\sc top}+{\rm UV}_{\sc stop}^{(1)}+{\rm UV}_{\sc stop}^{(2)}+{\rm UV}_{\sc 
stop}^{(3)}+{\rm UV}_{\sc stop}^{(4)}=0
$$
It is the main result of this section. Such a cancellation of divergences in relations 
originating from the supersymmetry is a typical feature of SUSY-theories 
\cite{Wess}. 
  
%
%
%
%
%
%
%
%
%
%
%
%
%
%
%
%
%
%
%
%
\section{Finite part of $\Delta$}

The topic of this section is to compute the finite part of expression (\ref{correction}). 
Having proved the total cancellation of divergences it can be easily shown that the finite 
part does not depend on the mass-scale $\mu$. This can be seen from the fact, that the 
only $\mu$-dependent factor 
$\ln \mu^2$ can be joined to the divergent factor $\div$ which drops out. In the 
explicit expressions we may, for convenience, put $\mu=1$ (mass unit). 

The idea of the following computation is to split the set of the considered diagrams with 
respect to the magnitude of the typical mass and discuss these clusters of contributions 
separately.
At this place it is indeed necessary to mention several restrictions put on the parameters 
of the theory:
\begin{itemize}
\item The masses of squarks  and the top
mass  are much
bigger than the other masses involved in the computation -
$m_\sto,m_\stt,m_t \gg m_W,m_A,m_H,m_h$)
\item The off-diagonal entries in the $\tilde{t}_L-\tilde{t}_R$
mass-matrix are very small with respect to $m_t^2$ which implies that  there is no 
significant
mixing in the supertop sector (see Appendix B for clarification). This condition can be 
translated in the mathematical form as $A_tm_6+\mu\cot{\beta}{} \ll m_t$.
\end{itemize}
The first condition is relevant for (almost) the whole
(experimentally admissible) area of the MSSM parametric
space with the only exceptional case that $m_A$ is very massive. However, the
mass $m_A$ often appears together with the factor $\cos{\beta}{}$,
which is expected to be small enough to suppress such term. 
The second condition is more speculative but seems to be
true for all the squark species (see \cite{Ber}); for our purposes 
this will be taken as an assumption . 

From the previous lines and the explicit form of the contributions listed in Appendix A it can 
be easily seen that the most important terms should be of order $m_{\tilde{t}}^2$. 
However, due to the previous assumptions, this term turns out to be small compared to the 
contribution 
coming from  the terms of order $\frac{m_t^4}{m_Z^2}$. 

The discussion below proves this statement.
The notation is again 
abbreviated as in the previous section i.e. there will be no difference between the symbols 
for the finite part and the full contribution of the examined diagram.

\subsection{Contributions of magnitude $m_{\tilde{t}_i}^2$}

Looking at  the coupling constants in Appendix A and taking into account the relation 
(\ref{stopmasssplit}) from Appendix B one can check that the only contributions 
proportional to 
$m_\sto^2 m_\stt^2$ come from the second, third and fourth type graphs in (\ref{stopsector}). 
The relevant expression is defined by
\beqa
{\rm F}_{m_{\tilde{t}_i}^2}&\equiv& 
\left[B_Z^{\tilde{t}_1\tilde{t}_2}+B_Z^{\tilde{t}_2\tilde{t}_2}+
B_Z^{\tilde{t}_1\tilde{t}_2}-T_A^{h\tilde{t}_1}-T_A^{h\tilde{t}_2}
-T_A^{H\tilde{t}_1}-T_A^{H\tilde{t}_2}+\right.\nonumber\\
&&\left.+S_h^{\tilde{t}_1}+S_h^{\tilde{t}_2}+S_H^{\tilde{t}_1}+S_H^{\tilde{t}_2}
-S_A^{\tilde{t}_1}-S_A^{\tilde{t}_2}+S_Z^{\tilde{t}_1}+S_Z^{\tilde{t}_2}\right]_{m_{\tilde{
t}_i}^2 \> only}
\eeqa
After some manipulations one obtains
\beqa
{\rm F}_{m_{\tilde{t}_i}^2}=\frac{N_cg^2m_Z^2}{16\pi^2m_W^2}
\Biggl\{
2\left({\vare^t_L}^2\cos{\tt}{2}+{\vare^t_R}^2\sin{\tt}{2}\right)
m_\sto^2\left(1-\ln{m_\sto^2} \right)+
\nonumber\\
+2
\left({\vare^t_L}^2\sin{\tt}{2}+{\vare^t_R}^2\cos{\tt}{2}\right)
m_\stt^2\left(1-\ln{m_\stt^2} \right)-
\nonumber\\
-\left({\vare^t_L}\cos{\tt}{2}+{\vare^t_R}\sin{\tt}{2}\right)^2
\left[
2m_\sto^2
-2\intx D_{m_Z}^{\sto\sto}(x)
\ln{D_{m_Z}^{\sto\sto}(x)}
\right]-
\nonumber\\
-\left({\vare^t_L}\sin{\tt}{2}+{\vare^t_R}\cos{\tt}{2}\right)^2
\left[
2m_\stt^2
-2\intx D_{m_Z}^{\stt\stt}(x)
\ln{D_{m_Z}^{\stt\stt}(x)}
\right]-\>\nonumber
\eeqa
$$
-\frac{1}{2}\sin{2\tt}{2}
\left({\vare^t_L}-{\vare^t_R}\right)^2
\left[
m_\sto^2+m_\stt^2
-2\intx D_{m_Z}^{\sto\stt}(x)
\ln{D_{m_Z}^{\sto\stt}(x)}
\right]\Biggr\}-
$$
\beq
\label{konecnacast}-\frac{N_cg^2m_t^2}{32\pi^2m_W^2\sin{\beta}{2}}
\left(A_tm_6\sin{\beta}{}+\mu\cos{\beta}{}\right)^2
\intx\ln{D_0^{\sto\stt}(x)}
\eeq
Note that the last term must indeed be taken into account here because of 
(\ref{stopmasssplit}). Fortunately it is strongly suppressed by the assumption of a small 
mixing in the supertop sector, see Appendix B.

First, it can be checked immediately that the terms of the form $const. \times 
m_{\tilde{t}_i}^2$ exactly cancel. The remaining structure is already not so easy to 
handle. 
The assumption of a small mixing allows us to approximate $\sin{\tt}{}\sim 0$, which 
drops out of the penultimate term in (\ref{konecnacast}). The key role of this observation 
consists in the fact that the rest of (\ref{konecnacast}) already does not involve any mixed 
$D_{m_Z}^{\sto\stt}(x)
$ term. Thus the expressions proportional to $m_\sto^2$ and $m_\stt^2$ split into  two 
independent clusters.  
To proceed one can use the  expansion
\beq
\label{expansion}\ln{D_{m_Z}^{\tilde{t}_i\tilde{t}_i}(x)}=
\ln{m_{\tilde{t}_i}^2}+\ln\left[1-\frac{m_Z^2}{m_{\tilde{t}_i}^2} x(1-x)\right]=
\ln{m_{\tilde{t}_i}^2}-\frac{m_Z^2}{m_{\tilde{t}_i}^2} x(1-x)+
O\left(\frac{m_Z^4}{m_{\tilde{t}_i}^4} \right)
\eeq
which originates from the definition of $D_{m_Z}^{\tilde{t}_i\tilde{t}_i}$, see Appendix A. 
Neglecting the contributions proportional to $m_Z^2$ one can check that the terms of the 
type $m_{\tilde{t}_i}^2\ln{m_{\tilde{t}_i}^2}$ cancel too. 
The previous discussion leads to the following result: 
\begin{itemize}
\item In the case of no significant mixing within the supertop sector the contribution of the 
magnitude $m_{\tilde{t}_i}^2$ turns out to be negligible compared to the correction 
proportional to $\frac{m_t^4}{m_Z^2}$, which is investigated in the next subsection. 
\end{itemize} 
If the mixing in the supertop sector is not negligible one can obtain 
large negative contribution proportional to $m_{\tilde{t}_i}^2$ from this cluster of 
diagrams; for more comprehensive discussion see \cite{Ber}.


\subsection{Contribution proportional to ${m_t^4}{m_Z^{-2}}$, leading $log$-term}

Looking into  (\ref{topsector}) and (\ref{stopsector}), 
one can immediately write down the sum of relevant terms :      
\beqa
{\rm F}_{m_t^4 \times M^{-2}}&\equiv& 
\biggl[B_h^t+B_H^t-B_A^t-T_A^{ht}-T_A^{Ht}+\nonumber\\
&&+B_h^{\sto\sto}+B_h^{\stt\stt}+B_H^{\sto\sto}+B_H^{\stt\stt}
\biggr]_{m_t^4 \times M^{-2} \> only}
\label{definice}
\eeqa
Note that the graph $B_Z^t$ produces only a factor of order $m_t^2$ and therefore does 
not belong into this sum.

Let us  first deal with the graphs involving top quark loop (the first five terms in 
(\ref{definice})). One can easily check that they contribute by
$$
{\rm F}_{m_t^4 \times M^{-2}}^{top}
=\frac{N_cg^2m_t^2}{16\pi^2m_W^2\sin{\beta}{2}}\times
$$
$$\times\Biggl\{
\cos{\alpha}{2}\left[m_t^2
+\intx\ln{D_{m_h}^{tt}(x)}\left[-3m_t^2+3m_h^2x(1-x)\right]
\right]+
$$
\beq
+\sin{\alpha}{2}
\left[m_t^2
+\intx\ln{D_{m_H}^{tt}(x)}\left[-3m_t^2+3m_H^2x(1-x)\right]
\right]+
\eeq
$$
+\cos{\beta}{2}\left[-m_t^2+
\intx\ln{D_{m_A}^{tt}(x)}\left[m_t^2-3m_A^2x(1-x)\right]
\right]
-m_t^2\sin{\beta}{2}
\left(1-\ln{m_t^2} \right)\Biggr\}
$$
Note that all the irrelevant parts of orders $m_t^2$, $m_Z^2$,$m_h^2$ and $m_H^2$ 
were neglected; the 
possibly large factor $\sim m_A^2m_t^2\times M^{-2}$ is suppressed by 
$\cos{\beta}{}$. The other factors like $\sim m_h^2m_t^2\times M^{-2}$ or $\sim 
m_H^2m_t^2\times M^{-2}$ are assumed not to be high above $\sim m_Z^2m_t^2\times 
M^{-2}$, moreover they are put down by an overall  factor coming from 
the integration over $x$.  The magnitude of the total error is approximately 10 \%. In 
addition it is easy to see that the non-logarithmic factors cancel. 

The situation in the supertop-loop cluster (corresponding to the last four terms in 
(\ref{definice})) is again quite complicated due to the structure of the appropriate squares 
of the couplings (\ref{konec1}). Extracting only the relevant parts 
involving $m_t^4$ one gets
$$
{\rm F}_{m_t^4 \times M^{-2}}^{stop}=
\frac{N_cg^2m_t^4}{16\pi^2m_W^2\sin{\beta}{2}}
\left\{
\cos{\alpha}{2}
\left[
\intx\ln{D_{m_h}^{\sto\sto}(x)}+
\intx\ln{D_{m_h}^{\stt\stt}(x)}
\right]+
\right.
$$
\beq
+
\left.
\sin{\alpha}{2}
\left[
\intx\ln{D_{m_H}^{\sto\sto}(x)}+
\intx\ln{D_{m_H}^{\stt\stt}(x)}
\right]
\right\}
\eeq
The last thing to be done is to apply expansion similar to (\ref{expansion}) on the 
$D$'s in the integrals. Summing up the particular results above it is straightforward 
to obtain the leading logarithmic term ($N_c=3$)
\beq
\label{nejdulezitejsi}\Delta=\frac{3g^2m_t^4}{16\pi^2m_W^2\sin{\beta}{2}}
\ln{\left(\frac{m_\sto^2m_\stt^2}{m_t^4}\right)}
+ \ldots
\eeq

This is the main result of the whole computation. It shows that
including the leading logarithmic one-loop correction
the original tree-level relation (\ref{rel1}) can be recast in
the form
$$
m_h^2+m_H^2-m_A^2-m_Z^2=\frac{3g^2m_t^4}{16\pi^2m_W^2\sin{\beta}{2}}
\ln{\left(\frac{m_\sto^2m_\stt^2}{m_t^4}\right)}+\ldots
$$
This relation agrees  with the
results presented in the literature (\cite{Ber}, \cite{GaT}).

Note that the relative error of the approximations used to derive the
previous relation does not exceed $10$\%. It is mainly due to neglecting all the
terms of order $m_t^2$ and lower. Next, the form of the leading
term (\ref{nejdulezitejsi}) is invalid in case that any significant
mixing in the stop sector occurs.

%
%
%
%
%
%
%
%
%
%
%
%
%
%
%
%
%
%
%
%
\section*{Conclusion}
The paper is devoted to  one of the most important features of the SUSY-theories -- 
the total cancellation of ultraviolet divergences in the relations originating from the 
supersymmetry. Although it is expected to be so in general, there is still no explicit 
proof based on the Ward identities. 
Therefore it is convenient to demonstrate this mechanism at least in a  particular case of the 
MSSM tree-level relation $\rel$. The rule is renormalised 
using the usual diagrammatic technique. The only considered one-loop graphs are 
those involving top and supertop loops because the expected magnitude of the 
correction is the largest (taking into account the observation of \cite{GaT} 
that no significant contribution 
descends from the chargino-neutralino sector). 
The original results of \cite{Ber} and \cite{YaY} are recovered performing 
the whole computation 
in the unitary gauge; the explicit mechanism of the divergence cancellation is shown. 
The finite part is discussed in detail in the case of no mixing in the supertop sector. 
%
%
%
%
%
%
%
%
%
\appendix
\section{ Relevant Feynman graphs}
%
%
%
%
%
%
%
%
%
%
This Appendix contains all the graphs discussed in previous
sections. They are divided into two main subgroups - diagrams with
quarks in loops and diagrams with the corresponding SUSY-partners. 
Each group consists of several types of
graphs; the notation is usual and (perhaps) self-explanatory.
The special symbols are defined as follows:
\begin{itemize}
\item $\div\equiv \vare^{-1}-\gamma_{\sc e}+\ln{4\pi} \>\>$
denotes the "divergent" part of a graph; here $2\vare=4-d$;
$d$ is the noninteger dimension used in the dimensional
regularisation procedure; $\gamma_{\sc e}$ is the
Euler-Mascheroni constant.
\item $B^{f_1f_2..}_{X}$, $T^{f_1f_2..}_{X}$ and  $S^{f_1f_2..}_{X}$
denote self-energies descending from the  Feynman graphs (usualy called \underline{b}lobs,
\underline{t}adpoles
and \underline{s}eagulls) with external lines $X$ and
internal $f_1,f_2,..$.
\item $D_{q}^{f_1f_2}(x) \equiv
m_{f_1}^2(1-x)+m_{f_2}^2x-q^2x(1-x)$ is the common factor arising
from the regularisation prescription (see \cite{Aok}) for
UV-divergent graphs; $q$ is the momentum of the incoming
(and outgoing) particle; in the on-shell scheme $q^2=m_X^2$.
\item The constants $g_{f_1f_2f_3..}$ denote the numerical parts of the
corresponding vertices . The non-number parts are
contained in the structure of integrands.
\item For the sake of brevity the overall colour factor $N_c=3$ is suppressed  but must be 
included to obtain the correct results.
\end{itemize}
\subsection{\label{topsector}Graphs with top quark loops}
\subsection*{I. Scalar and pseudoscalar self-energies:
}
$$
\qquad
\parbox{30mm}
{
\epsffile{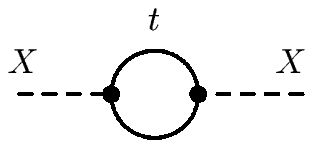}
}
\qquad\equiv \qquad
B^{t}_X(q);\qquad X\equiv h,H,A
$$
\beqa
B^{t}_X(q)&=&-ig^2_{Xtt}\mu^{2\vare}\intdp{k}{
\tr\frac{i}{\s{k}-m_t}\frac{i}{\s{k}-\s{q}-m_t}};\qquad X=h,H
\nonumber\\
B^{t}_A(q)&=&-ig^2_{Xtt}\mu^{2\vare}\intdp{k}{
\tr\gamma_5\frac{i}{\s{k}-m_t}\gamma_5\frac{i}{\s{k}-\s{q}-m_t}}
\nonumber
\eeqa
Using the routine calculational procedure (see for instance \cite{Aok}) the
results are
\beqa
B^{t}_X(q)&=&g^2_{Xtt}
\frac{1}{4\pi^2}\left\{
\div\left(3m_t^2-\frac{q^2}{2}\right)+\right.
\left(m_t^2-\frac{q^2}{6}\right)+\nonumber\\
&+&
\left.\intx\mln{D^{tt}_q(x)}\left[-3m_t^2+3q^2x(1-x)\right]
\right\};\qquad X=h,H
\\
B^{t}_A(q)&=&g^2_{Att}
\frac{1}{4\pi^2}\left\{
\div\left(-m_t^2+\frac{q^2}{2}\right)-\right.
\left(m_t^2-\frac{q^2}{6}\right)+\nonumber\\
&+&
\left.\intx\mln{D^{tt}_q(x)}\left[m_t^2-3q^2x(1-x)\right]
\right\}\nonumber
\eeqa
The corresponding couplings are
\beq
\label{konstanty}
g_{htt}= -\frac{igm_t\cos{\alpha}{}}{2m_W\sin{\beta}{}}; \qquad
g_{Htt}= -\frac{igm_t\sin{\alpha}{}}{2m_W\sin{\beta}{}}; \qquad
g_{Att}= -\frac{gm_t}{2m_W}\cot{\beta}{}
\eeq
\subsection*{
II. Z-boson self-energy graph (vacuum polarisation
tensor):}
$$
\qquad
\parbox{30mm}
{
\epsffile{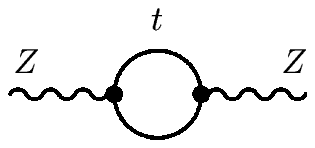}
}
\qquad\equiv \qquad
B^{t}_Z(q)^{\mu\nu}
$$
$$
B^{t}_Z(q)^{\mu\nu}=-ig^2_{Ztt}\mu^{2\vare}\intdp{k}{
\tr\frac{i\gamma^\mu K(\gamma)}{\s{k}-m_t}
\frac{i\gamma^\nu K(\gamma)}{\s{k}-\s{q}-m_t}};
$$
where
$$
K(\gamma)\equiv ({\vare_L^t}+{\vare_R^t})I_4-({\vare_L^t}-{\vare_R^t})\gamma_5.$$
The constants ${\vare_L^t}$, ${\vare_R^t}$ 
are defined as follows (in general $\vare^f=T_3-Q_f\sin{\tW}{2}$)
\beq
{\vare_L^t}\equiv\frac{1}{2}-q_t\sin{\tW}{2}; \qquad {\vare_R^t}\equiv -q_t\sin{\tW}{2}
\eeq
The coupling $g_{Ztt}$ is
\beq
g_{Ztt}= -\frac{ig}{2\cos{\tW}{}}
\eeq
Performing the usual steps the result becomes
\beqa
B_Z^{t}(q)^{\mu\nu}& = &
g^2_{Ztt}
\left\{
-\frac{1}{4\pi^2}\gmn\!\!\intx\!\mln{D_q^{tt}(x)}
\left(-\frac{m_t^2}{2}\right)\right.
\nonumber \\
&+&\frac{1}{4\pi^2}\div
\left[-\frac{m_t^2}{2}\gmn +
\frac{2}{3}({\vare_L^t}^2+{\vare_R^t}^2)(\gmn q^2-q^\mu
q^\nu)\right]+\\
&-&\left.\frac{1}{2\pi^2}\intx\mln{D_q^{tt}(x)}
2({\vare_L^t}^2+{\vare_R^t}^2)(q^\mu q^\nu-\gmn q^2)x(1-x)\right\}\nonumber
\eeqa
\subsection*{
III. Tadpoles involving top quark loop:}
In general there are eight graphs to be considered in this
paragraph. They are the following 
(here $X=h$,$H$,$A$ and $S=h$,$H$):
\vskip 2mm
$$
\parbox{30mm}
{
\epsffile{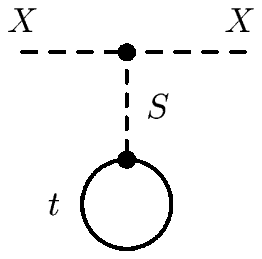}
}
\equiv \qquad
T^{St}_X(q)\qquad
\parbox{30mm}
{
\epsffile{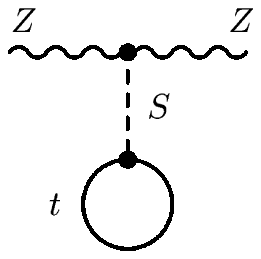}
}\equiv \qquad
T^{St}_Z(q)^{\mu\nu}
$$
Fortunately many of them cancel because of a nice
property of the corresponding couplings "sitting" in the upper
vertex
\beqa
\label{relfajn1}\label{relfajn2}
g_{hhh}+g_{HHh}+g_{ZZh}=0 \nonumber \\
g_{hhH}+g_{HHH}+g_{ZZH}=0
\eeqa
The remaining graphs are evaluated in the usual way:
$$
T^{St}_A(q)=ig_{AAS}g_{ttS}\frac{i}{m^2_S}\mu^{2\vare}\intdp{k}{
\tr\frac{i}{\s{k}-m_t}};
$$
This d-dimensional integration is already easy to handle; we get
\beq
T^{St}_A(q)=4g_{AAS}g_{ttS}\frac{m_t^3}{m^2_S}\frac{1}{16\pi^2}
\left(\div+1-\mln{m^2_t}\right);
\eeq
The couplings $g_{htt}$ and $g_{htt}$ are already
written in (\ref{konstanty}); the remaining constants are
\beq
\label{konstanty2}
g_{AAh}= -\frac{igm_Z}{2\cos{\tW}{}}\cos{2\beta}{}\sin{(\alpha+\beta)}{}; \qquad
g_{AAH}= \frac{igm_Z}{2\cos{\tW}{}}\cos{2\beta}{}\cos{(\alpha+\beta)}{}
\eeq

At the end of this subsection note that the coupling constants used
in this article can be found for example
in \cite{GaH} and \cite{GaT}. In the case of
supertops one must transform the rules in \cite{GaH} from the
$L-R$ basis to the supertop mass-diagonal basis $1-2$; this
procedure is well described in the cited paper.

 \subsection{\label{stopsector}Graphs with supertop loops}
\subsection*{
I. Diagrams of the first type:}
There are 7 graphs to be investigated in this cathegory, namely
$$
\qquad
\parbox{30mm}
{
\epsffile{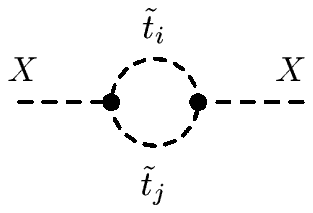}
}
\qquad\equiv \qquad
B^{\tilde{t}_i\tilde{t}_j}_X(q);\qquad X\equiv h,H,A;\qquad
i,j=1,2 \nonumber
$$
$$
B^{\tilde{t}_i\tilde{t}_j}_X(q)=
ig_{X\tilde{t}_i\tilde{t}_j}g_{X\tilde{t}_j\tilde{t}_i} \mu^{2\vare}\intdp{k}
\frac{i}{k^2-m_{\tilde{t}_i}^2}
\frac{i}{(k-q)^2-m_{\tilde{t}_j}^2}(2-\delta_{ij})
$$
Using again \cite{Aok}, the last expression can be simplified to the final form
\beq
B^{\tilde{t}_i\tilde{t}_j}_X(q)=g_{X\tilde{t}_i\tilde{t}_j}
g_{X\tilde{t}_j\tilde{t}_i}(2-\delta_{ij})
\frac{1}{16\pi^2}\left[\div-\intx\mln{D_q^{\tilde{t}_i\tilde{t}_j}(x)}\right]
\eeq
Note that the factor $(2-\delta_{ij})$ counts the number of
nonequivalent contractions.
To finish this paragraph it is necessary to specify the
couplings; the vertices involving scalars are symmetric with
respect to $i\leftrightarrow j$
\beqa
g_{ h\sto\sto}&=&
\frac{igm_Z}{\cos{\tW}{}}\sin{(\alpha+\beta)}{}
\left({\vare_L^t}\cos{\tt}{2}-{\vare_R^t}\sin{\tt}{2}\right)
-\frac{igm_t^2\cos{\alpha}{}}{m_W\sin{\beta}{}} \nonumber\\
& &-\frac{igm_t\sin{2\tt}{}}{2m_W\sin{\beta}{}}
\left(A_tm_6\cos{\alpha}{}-\mu\sin{\alpha}{}\right)
\nonumber
\\
g_{ h\stt\stt}&=&
\frac{igm_Z}{\cos{\tW}{}}\sin{(\alpha+\beta)}{}
\left({\vare_L^t}\sin{\tt}{2}-{\vare_R^t}\cos{\tt}{2}\right)
-\frac{igm_t^2\cos{\alpha}{}}{m_W\sin{\beta}{}}
\nonumber \\
& &+\frac{igm_t\sin{2\tt}{}}{2m_W\sin{\beta}{}}
\left(A_tm_6\cos{\alpha}{}-\mu\sin{\alpha}{}\right)
\nonumber\\
g_{ h\sto\stt}&=&
-\frac{igm_Z}{\cos{\tW}{}}\sin{(\alpha+\beta)}{}
\left({\vare_L^t}+{\vare_R^t}\right)\sin{\tt}{}\cos{\tt}{}
\nonumber \\
& &-\frac{igm_t\cos{2\tt}{}}{2m_W\sin{\beta}{}}
\left(A_tm_6\cos{\alpha}{}-\mu\sin{\alpha}{}\right)
\label{konec1}\\
g_{ H\sto\sto}&=&
-\frac{igm_Z}{\cos{\tW}{}}\cos{(\alpha+\beta)}{}
\left({\vare_L^t}\cos{\tt}{2}-{\vare_R^t}\sin{\tt}{2}\right)
-\frac{igm_t^2\sin{\alpha}{}}{m_W\sin{\beta}{}}
\nonumber \\
& &-\frac{igm_t\sin{2\tt}{}}{2m_W\sin{\beta}{}}
\left(A_tm_6\sin{\alpha}{}+\mu\cos{\alpha}{}\right)
\nonumber\\
g_{ H\stt\stt}&=&
-\frac{igm_Z}{\cos{\tW}{}}\cos{(\alpha+\beta)}{}
\left({\vare_L^t}\sin{\tt}{2}-{\vare_R^t}\cos{\tt}{2}\right)
-\frac{igm_t^2\sin{\alpha}{}}{m_W\sin{\beta}{}}
\label{H22} \nonumber\\
& &+\frac{igm_t\sin{2\tt}{}}{2m_W\sin{\beta}{}}
\left(A_tm_6\sin{\alpha}{}+\mu\cos{\alpha}{}\right)
\nonumber\\
g_{ H\sto\stt}&=&
\frac{igm_Z}{\cos{\tW}{}}\cos{(\alpha+\beta)}{}
\left({\vare_L^t}+{\vare_R^t}\right)\sin{\tt}{}\cos{\tt}{}
\nonumber \\
\label{H12}
& &-\frac{igm_t\sin{2\tt}{}}{2m_W\sin{\beta}{}}
\left(A_tm_6\sin{\alpha}{}+\mu\cos{\alpha}{}\right)
\nonumber\eeqa
while vertices with pseudoscalar $A$ are antisymmetric:
\beqa
g_{A\sto\stt}=-g_{A\stt\sto} &=&
\frac{gm_t}{2m_W\sin{\beta}{}}
\left(A_tm_6\cos{\beta}{}-\mu\sin{\beta}{}\right)
\nonumber\\
g_{A\sto\sto}=-g_{A\stt\stt} &=& 0
\eeqa
\subsection*{
II. Z-boson self-energy graphs with looping
superquarks:}
The graphs relevant to this paragraph are all of the type
$$
\qquad
\parbox{30mm}
{
\epsffile{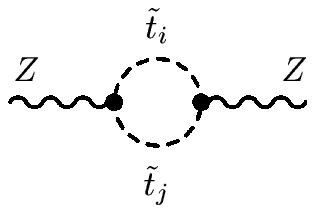}
}
\qquad\equiv \qquad
B^{\tilde{t}_i\tilde{t}_j}_Z(q)^{\mu\nu};\qquad j=1,2 \nonumber
$$
$$
B^{\tilde{t}_i\tilde{t}_j}_Z(q)^{\mu\nu}=
ig_{Z\tilde{t}_i\tilde{t}_j}^2
\mu^{2\vare}\intdp{k} \frac{i(2k-p)^\mu}{k^2-m_\sto^2}
\frac{i(2k-p)^\nu}{(k-q)^2-m_\sto^2}
$$
As in the previous cases, after some algebra one obtains
$$
B^{\tilde{t}_i\tilde{t}_j}_Z(q)^{\mu\nu}=
-g_{Z\tilde{t}_i\tilde{t}_j}^2\frac{1}{16\pi^2}\left\{
\div\left[\frac{1}{3}\left(q^\mu q^\nu-\gmn q^2\right)+
(m^2_{\tilde{t}_i}+m^2_{\tilde{t}_j})\gmn \right]\right.+
$$
$$
+\!\left.\gmn \left(m^2_{\tilde{t}_i}+m^2_{\tilde{t}_j}-\frac{q^2}{3}\right)\!-\!
\intx\left[q^\mu q^\nu(1-2x)^2+2\gmn D_q^{{\tilde{t}_i}{\tilde{t}_j}}(x)\right]
\mln{D_q^{{\tilde{t}_i}{\tilde{t}_j}}(x)}
\right\}
$$
\subsection*{
III. Tadpoles involving supertop loop:}
In general there is again many diagrams belonging to this paragraph; 
as in the previous section  the relations (\ref{relfajn1})-(\ref{relfajn2}) 
ensure that most of the graphs cancel. The remaining are
(here $S=h$,$H$ and $i=1,2$):
\vskip 2mm
$$
\parbox{30mm}
{
\epsffile{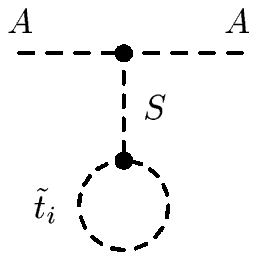}
}
\equiv \qquad
T^{S\tilde{t}_i}_A(q)
$$
The contributions coming from these graphs are 
$$
T^{S\tilde{t}_i}_A(q)=
-ig_{AAS}g_{\tilde{t}_i\tilde{t}_iS}\frac{i}{m^2_S}\mu^{2\vare}\intdp{k}\frac{i}{k^2-m_{\tilde{t}_i}^2};
$$
which gives after regularisation
\beq
T^{S\tilde{t}_i}_A(q)=
-g_{AAS}g_{\tilde{t}_i\tilde{t}_iS}\frac{m^2_{\tilde{t}_i}}{m^2_S}\frac{1}{16\pi^2}
\left(\div+1-\mln{m^2_{\tilde{t}_i}}\right);
\eeq
The necessary coupling constants are written in (\ref{konstanty2}) and (\ref{konec1})
\subsection*{
IV. Seagull graphs:} Due to the presence of
 quadrilinear vertices involving two Higgses and two superquarks there is an additional 
 sort of graphs in this section -- the so-called seagull graphs which look as ($S=h$,$H$,$A$ and $i=1,2$)
$$
\parbox{30mm}
{
\epsffile{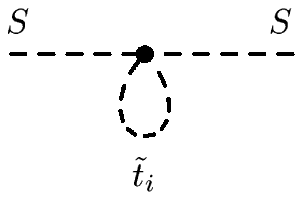}
}
\quad\equiv\quad S_S^{\tilde{t}_i}(q);\qquad
\parbox{30mm}
{
\epsffile{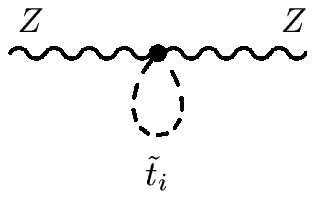}
}
\quad\equiv\quad S_Z^{\tilde{t}_i}(q)^{\mu\nu}
$$
\vskip 4mm
Note that similar graphs can not appear in the fermion sector because the quadrilinear  vertices 
involving fermions have mass dimensions $>4$ and they would  cause nonrenormalisability 
of the theory. The contributions originating from these graphs can be written as
$$
S^{\tilde{t}_i}_S(q)=
ig_{SS\tilde{t}_i\tilde{t}_i}\mu^{2\vare}\intdp{k}\frac{i}{k^2-m_{\tilde{t}_i}^2};
$$
$$
S^{\tilde{t}_i}_Z(q)^{\mu\nu}=
ig_{ZZ\tilde{t}_i\tilde{t}_i}\mu^{2\vare}\gmn\intdp{k}\frac{i}{k^2-m_{\tilde{t}_i}^2};
$$
which after regularisation gives
\beq
S^{\tilde{t}_i}_S(q)=
-g_{SS\tilde{t}_i\tilde{t}_i}m^2_{\tilde{t}_i}\frac{1}{16\pi^2}
\left(\div+1-\mln{m^2_{\tilde{t}_i}}\right);
\eeq
$$
S^{\tilde{t}_i}_S(q)^{\mu\nu}=
-g_{ZZ\tilde{t}_i\tilde{t}_i}m^2_{\tilde{t}_i}\gmn\frac{1}{16\pi^2}
\left(\div+1-\mln{m^2_{\tilde{t}_i}}\right);
$$
In general there are 8 graphs to deal with. Fortunately the coupling constants can be nicely summed up so that
\beq
g_{hh\sto\sto}+g_{HH\sto\sto}+
g_{hh\stt\stt}+g_{HH\stt\stt}-g_{AA\sto\sto}-g_{AA\stt\stt}=g_{GG\stt\stt}+g_{GG\sto\sto}
\eeq
where
\beqa
g_{GG\sto\sto}&=&
-\frac{ig^2}{2\cos{\tW}{2}}\cos{2\beta}{}
\left({\vare_L^t}\cos{\tt}{2}-{\vare_R^t}\sin{\tt}{2}\right)
-\frac{ig^2m_t^2}{2m_W^2}
\nonumber \\
g_{GG\stt\stt}&=&
-\frac{ig^2}{2\cos{\tW}{2}}\cos{2\beta}{}
\left({\vare_L^t}\sin{\tt}{2}-{\vare_R^t}\cos{\tt}{2}\right)
-\frac{ig^2m_t^2}{2m_W^2}
\eeqa
Note that these constants are exactly the couplings of the would-be Goldstone boson 
(which is within U-gauge absent). The last unspecified parameters are the couplings 
$g_{ZZ\tilde{t}_i\tilde{t}_i}$ :
\beqa
g_{ ZZ\sto\sto}&=&
\frac{2ig^2}{\cos{\tW}{2}}
\left({\vare_L^t}^2\cos{\tt}{2}+{\vare_R^t}^2\sin{\tt}{2}\right)
\nonumber \\
g_{ ZZ\stt\stt}&=&
\frac{2ig^2}{\cos{\tW}{2}}
\left({\vare_L^t}^2\sin{\tt}{2}+{\vare_R^t}^2\cos{\tt}{2}\right)
\eeqa
%
%
%
%
%
%
%
%
%
%
%
%
%
%
%
%
%
%
%
%
\section{Some useful  relations and comments}
This Appendix is devoted to several clarifications necessary to make the text more 
self-consistent. 

The first note refers to the MSSM itself. The full-range discussion of the relevant part of the 
MSSM classical
 lagrangian is obviously out of the scope of this paper. There are several comprehensive 
 works in the literature
 that can be used for this purpose, namely \cite{GaH}, \cite{Sim}, \cite{Bailin} or 
 \cite{Wess}. The notation
 is similar to that used in \cite{GaH}. 

The rest of the Apppendix contains some comments on technical details of the computation. 
To enable the
 reader follow
 the steps described above it is necessary to write down several not so well known
 tree-relations often used
 during the computation. First of them,
\beq
m_h^2\cos{\alpha}{2}+m_H^2\sin{\alpha}{2}
-m_A^2\cos{\beta}{2}-m_Z^2\sin{\beta}{2}=0
\eeq
can be derived by utilising (\ref{rel1}) and relations
\beq
\sin{2\alpha}{}=
\left(\frac{m_h^2+m_H^2}{m_h^2-m_H^2}\right)\sin{2\beta}{};\qquad
\cos{2\alpha}{}=
\left(\frac{m_A^2-m_Z^2}{m_h^2-m_H^2}\right)\cos{2\beta}{}
\label{hmotyhiggsu}
\eeq
that can be found for example in \cite{Sim}. Note only that the parameters $\alpha$ 
and $\beta$ are the mixing angles in the scalar and pseudoscalar parts of the Higgs sector. 
These relations are also very handy if we want to express (tree) Higgs masses in terms of 
$m_Z^2$, $\alpha$ and $\beta$ dealing with the factors $m_h^{-2}$ or $m_H^{-2}$ 
coming from the tadpoles in (\ref{topsector}) and (\ref{stopsector}). 

The next thing to be clarified is the role of the parameters $A_tm_6$ and $\mu$ in the 
supertop mass-squared matrix. This matrix in the $L-R$ basis  looks (see \cite{Ber})
\beq
\label{matice}
M^2_{\tilde{t}_{L,R}}=\left(
\begin{array}{cc}
A & B \\
B & C
\end{array}
\right)
\qquad
\eeq
{\rm where}
\beqa
A & = & M^2_{\sc q}+m_Z^2\cos{2\beta}{}{\vare_L^t}+m_t^2 \nonumber\\
B & = & m_t\left(A_tm_6+\mu\cot{\beta}{}\right) \nonumber \\
C & = & M^2_{\sc u}+m_Z^2\cos{2\beta}{}{\vare_R^t}+m_t^2 \nonumber
\eeqa
is the usual parametrisation of its entries. (Note that the constants 
$M^2_{\sc q}$ and $M^2_{\sc u}$ are the so-called soft-SUSY breaking terms
which in general split the masses of supertops and shift them high above $m_t$, \cite{GaH}.)
The eigenvalues of this matrix can be easily derived in the form
\beq
\label{eigen}
m_{\tilde{t}_{1,2}}^2=\frac{1}{2}\left[A+C\pm\sqrt{(A+C)^2-4(AC-B^2)}\right]
\eeq
The mixing (diagonalising) angle $\tt$ is then defined by 
$$
\tan{\tt}{}=\frac{2B}{A-C}
$$
which can be rewritten in terms of the eigenvalues (\ref{eigen}) as follows
\beq
\label{stopmasssplit}\sin{2\tt}{}=
\frac{2m_t(A_tm_6+\mu\cot{\beta}{})}{m_\sto^2-m_\stt^2}
\eeq
This relation connects the off-diagonal entries in the supertop mass-squared matrix with the 
magnitude of the
 supertop mass-split and the mixing angle $\tt$. Assuming now that the supertop mass-squared 
 does not
 exceed  the top-scale too much the assumption of a small mixing can be recast in the form 
 $(A_tm_6+\mu\cot{\beta}{})\ll m_t$.
%
%
%
%
%
%
%
%
%
%
%

\bigskip
{\small I would like to thank to my supervisor Prof. Ji\v{r}\'{\i} Ho\v{r}ej\v{s}\'{\i} for all
the comments and helpful suggestions during the long period of writing this article. 
The work has been partially supported by the grant GA\v{C}R No. 202/98/0506. 
}

\end{document}